\documentclass[prd,twocolumn,groupedaddress,nofootinbib,preprintnumbers,nofootinbib,superscriptaddress]{revtex4} 

\usepackage{epsfig,graphicx}
\usepackage{dcolumn}% Align table columns on decimal point
\usepackage{bm}% bold math
\usepackage{latexsym}
\usepackage{amsfonts}
\usepackage[usenames,dvipsnames]{color}
\usepackage{enumerate}
\usepackage{amssymb}
\usepackage{amsmath}

\begin{document}

\title{Topological Inflation with Large Tensor-to-scalar Ratio}

\preprint{IPMU14-0087}

\author{Yu-Chieh Chung}
\email{yuchieh.chung@ipmu.jp}
\author{Chunshan Lin}

\email{chunshan.lin@ipmu.jp}
\affiliation{Kavli Institute for the Physics and Mathematics of the Universe (WPI), Todai Institutes for Advanced Study, University of Tokyo, 5-1-5 Kashiwanoha, Kashiwa, Chiba 277-8583, Japan}

\date{\today}% It is always \today, today,

\begin{abstract}

BICEP2's detection on the primordial B-mode of CMB polarization suggests that inflation occurred around GUT scale, with the tensor-to-scalar ratio $r\simeq0.2$. Inspired by this discosvery, we study the topological inflation which was driven by a double/single/no well potential. We show that with proper choice of parameters, all these three types of topological inflationary models could be consistent with the constraints from current observations.

%with Higgs type of double well potential and axion-like potential, topological inflation has to happen near the edge of soliton, to provide the proper spectral tilt and tensor-to-scalar ratio. However, with  proper choice of inflationary potential, such model could be consistent with current observations. 
\end{abstract}
\maketitle

\section{Introduction}

Inflationary paradigm \cite{guth1980} has become a leading scenario of the early universe. It provides a very convincing solution to the flatness problem, horizon problem, and monopole problem in standard hot big bang cosmology.  It is also believed that the quantum fluctuation during inflation seeds the large scale structure and CMB anisotropies nowadays.

Inflation is successful but, nonetheless, also encounters some questioning on its validity \cite{Brandenberger:2012uj}. One of the conceptual problems is the fine-tuning.  Inflation was introduced to eliminate the fine-tuning condition for the initial data set of cosmology, i.e. the so called horizon problem and flatness problem. However, inflation itself requires fine-tuning on the initial conditon, which renders the fine-tuning problem returns in a different guise. 

In the context of topological inflation \cite{Linde:1994hy}\cite{Vilenkin:1994pv}, such fine-tuning could be alleviated. Topological inflation is a wild class of models where the inflaton field is forced to stay near a local  maximum of potential for topological reasons. 

Generally, a topological defect forms during the phase transition with spontaneous symmetry breaking. Its existence is related to the topology of the boundary of space, the topology of the set of vacua, and the existence of a nontrivial map from the boundary of space to the set of vacua. During the early universe, if the size of the soliton greater than the Hubble radius, inflation occures at the core of such a topological defect \cite{Linde:1994hy}\cite{Vilenkin:1994pv}.

The recent BICEP \cite{Ade:2014gua}\cite{Ade:2014xna} measurement of primordial B mode in the polarization of cosmic microwaves background suggests that inflation occured at the energy scale of $10^{16}$ GeV, with the tensor-to-scalar ratio $r\simeq0.2$.  This discovery definitively affects our understanding of early universe, see the following up research after BICEP2\cite{paperflood}. Lots of models would be ruled out due to the discovery of primordial B-mode.  In this paper we check examples of the topological inflation by using recent BCEP and established cosmological results.

The rest of the paper is organized as follows: at section II, we will check the consistency with experiment on the  well-known double-well potential model; at section III, we will check the case with single-well model, and at section IV, we will check the model without any well. We conclude and summarize our results at the final section. 

\section{topological inflation with double well potential}
%In this section, we briefly review the basic idea of topological inflation \cite{Vilenkin:1994pv}.

The most often considered potentials are Higgs type of $\phi^4$ potential and axion-like potential. 
Let's start from a Higgs type of double well potential, such as
\begin{eqnarray}
V(\phi)=\frac{1}{4}\lambda\left(\phi^2-\upsilon^2\right)^2~.
\end{eqnarray}
See \cite{higgscosmo} for an early study on this potential in the context of inflation. At the early universe, as universe expands, the phase transition splits space into two domains with $\phi=\upsilon$ and $\phi=-\upsilon$. Inflation happens in the core of domain wall if the thickness of the wall is larger than the Hubble radius, which requires $\upsilon>M_p$. Such largeness of $\upsilon$ implies that the effective mass of scalar field $m_{\phi}^2=\lambda\upsilon^2<<H^2$ (with help of Freedman equation), and scalar field undergo a period of slow-rollover. The virtue of this model, of which we are free from fine tuning the initial conditions, is attractive for theoretical study.

The slow roll parameters can be calculated as,
\begin{eqnarray}\label{slpara}
\epsilon&=&\frac{1}{2}\left(\frac{M_pV'}{V}\right)^2=\frac{8 \phi ^2 M_p^2}{\left(\upsilon ^2-\phi ^2\right)^2}~,\nonumber\\
\eta&=&\frac{M_p^2V''}{V}=-\frac{4 M_p^2 \left(\upsilon ^2-3 \phi ^2\right)}{\left(\upsilon ^2-\phi ^2\right)^2},\nonumber\\
n_s-1&=&-\frac{8 M_p^2 \left(\upsilon ^2+3 \phi ^2\right)}{\left(\upsilon ^2-\phi ^2\right)^2}~,\nonumber\\
r&=&16\epsilon~.
\end{eqnarray}
The spectral tilt $n_s-1$ and tensor-to-scalar ratio $r$ are experimental observables. Scalar tilt $n_s=0.9603\pm0.0073$ was given by PLANCK mission \cite{Ade:2013zuv}, and $r=0.2_{-0.05}^{+0.07}$ was given by BICEP2.  However, please notice that  $r=0.2$ is only consistent with the case of negligible foreground dusts. Considering the possiblity that this value could be contaminated by the non-negligible dust polarization signal, which was questioned in ref.\cite{Flauger:2014qra}, we choose a slightly smaller value
for tensor-to-scalar ratio\footnote{We would like to thank the anonymous referee for the suggestion on this point.}  $r=0.16_{-0.05}^{+0.06}$ that  was also suggested by BICEP2 \cite{Ade:2014gua}.
%of which BICEP2 team didn't clearly subtact, 

In addition to these two constraints, another constraint comes from the e-folding number which under slow roll approximation could be calculated as 
\begin{eqnarray}
N=\int_{\phi_e}^{\phi_i}\left(\frac{V}{M_p^2V'}\right)d\phi~,
\end{eqnarray}
where $\phi_i$ is the field value at the begin of inflationary period which corresponds to our largest scale physics nowadays,  and  $\phi_e$ denotes the one at the end of inflation, which in this case could be chosen as $\upsilon$ approximately.  

\begin{figure}
\begin{center}
\includegraphics[width=0.4\textwidth]{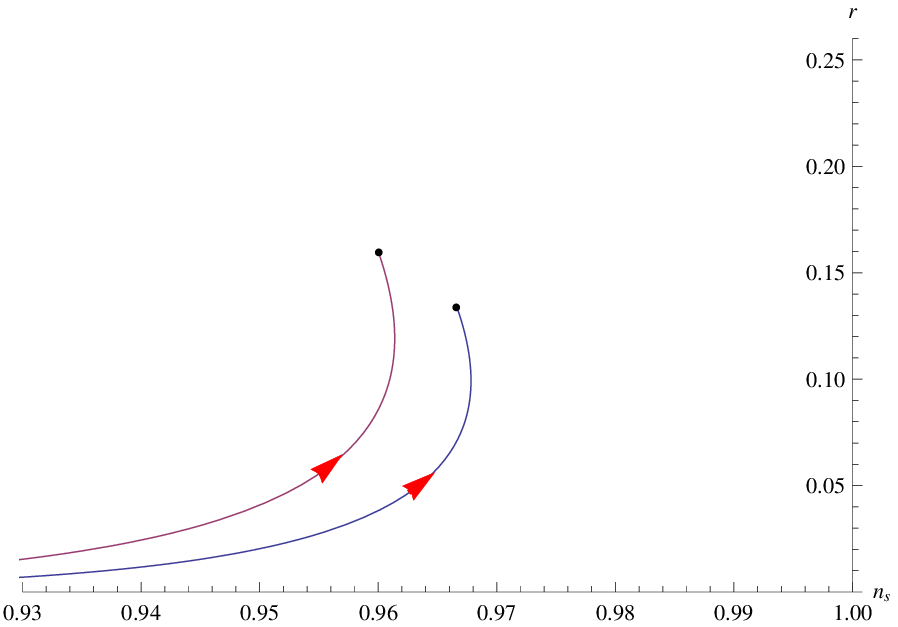}
\end{center}
\caption{$(n_s,r)$ curve in Higgs type of topological inflation. The blue curve corresponds to the inflation with 60 e-folding number, and the red curve corresponds to the one with 50 e-folding number. The red arrow denotes the direction of increasing $\phi/\upsilon$. The end points of  curves correspond to the vaccum $\phi=\upsilon$. }
\label{higgs}
\end{figure}

The $(n_s,r)$ curve is plotted  in FIG.\ref{higgs}. By comparing to the 95\% confidence contour of BICEP2 \cite{Ade:2014gua}, we can see that observation favors that inflation happens at somewhere around $\phi>\frac{3}{4}\upsilon$, which is near the edge of soliton. $\upsilon\gtrsim 50 M_p$ is required to ensure that inflation takes enough e-folding number.

\begin{figure}
\begin{center}
\includegraphics[width=0.4\textwidth]{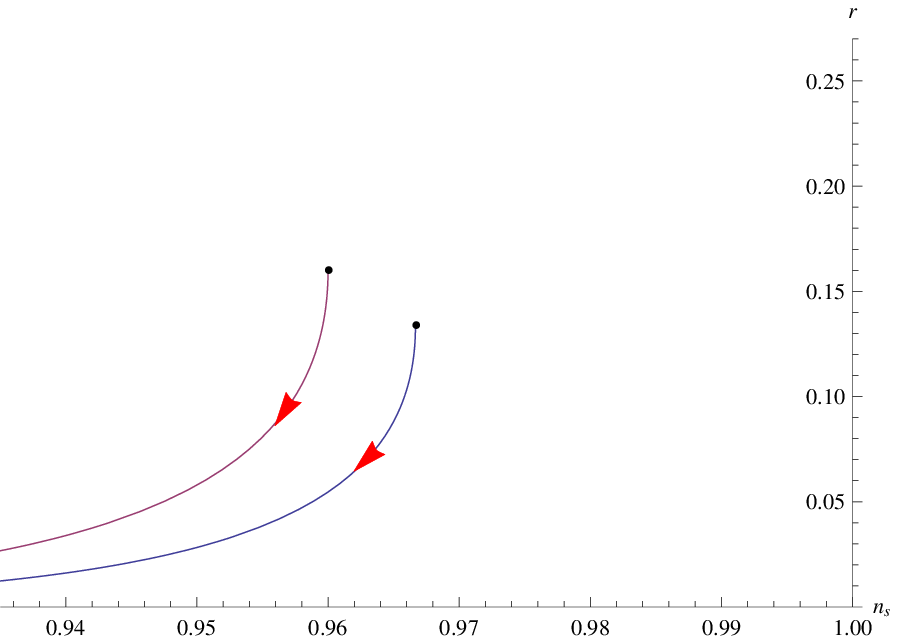}
\end{center}
\caption{$(n_s,r)$ curve in axion type of topological inflation. The blue curve corresponds to the inflation with 60 e-folding number, and the red curve corresponds to the one with 50 e-folding number. The red arrow denotes the direction of increasing $\phi/\upsilon$. The end points of curves correpond to the vacuum $\phi=0$. }
\label{axion}
\end{figure}

Then, let us consider the following axion-like potential\footnote{Although there are infinite number of wells in this potential, but only 2 wells are relevant to our study. Thus we still treat it as double-well model. }

\begin{eqnarray}
V(\phi)=\frac{m^2\upsilon^2}{n^2}\left[1-\cos(n\phi/\upsilon)\right]~,
\end{eqnarray} 
where $m$ is a mass scale and $n$ is an integer. Let's assume that our Hubble volume locates at somewhere between two vacua $\phi=0$ and $\phi=\frac{2\pi\upsilon}{n}$. 
 The slow roll parameters can be calculated as follows, 
\begin{eqnarray}
\epsilon&=&\frac{n^2 M_p^2 }{2 \upsilon ^2}\cdot \cot ^2\left(\frac{n \phi }{2 \upsilon }\right)~,\nonumber\\
\eta&=&\frac{n^2 M_p^2 }{2 \upsilon ^2}\cdot \cos \left(\frac{n \phi }{\upsilon }\right) \csc ^2\left(\frac{n \phi }{2 \upsilon }\right),\nonumber\\
n_s-1&=&\frac{n^2 M_p^2 }{ \upsilon ^2}\cdot\left[1-2\csc^2\left(\frac{n\phi}{2\upsilon}\right)\right],\nonumber\\
r&=&16\epsilon~.
\end{eqnarray}
We plot the $(n_s,r)$ curve in FIG.\ref{axion}, again, by comparing with results of BICEP2 \cite{Ade:2014gua}, we can see that the observation favors that inflation happens at somewhere $\frac{n\phi}{\upsilon}<\frac{\pi}{3}$, and $\upsilon\gtrsim 15M_p$ is required for enough e-folding number.

\section{Topological inflation with single well potential}

\begin{figure}
\begin{center}
\includegraphics[width=0.4\textwidth]{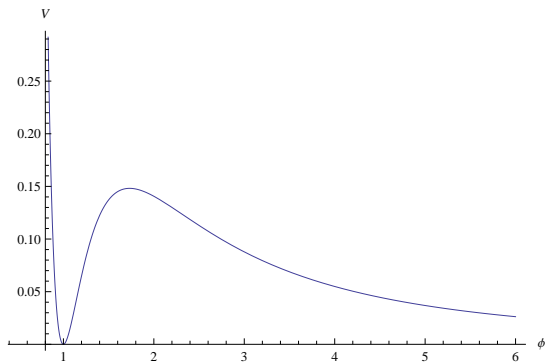}
\end{center}
\caption{Schematic plot of single well scalar potential with $\xi=\upsilon=1$.}
\label{potential}
\end{figure}

Now let us turn to check some  topological inflation models other than double well potential.  As an example, let us consider the following  potential motivated by extra-dimension compatification \cite{Denef:2007pq}
\begin{eqnarray}
V(\phi)=\xi\frac{\upsilon^6}{\phi^6}\left(\phi^2-\upsilon^2\right)^2~,
\end{eqnarray}
where $\upsilon$ is just a constant parameter with mass dimension, and $\xi$ is dimensionless parameter. See \cite{Ellis:1998gf} for an earlier work on single well topological inflation. This potential has two local minima, one is at $\phi=\upsilon$, the other is at infinite $\phi\to\infty$. It also has one local maximum, which is at $\phi=\sqrt{3}\upsilon$ (see FIG.\ref{potential}). %This potential is similar to the one of extra dimension compatification. 

\begin{figure}
\begin{center}
\includegraphics[width=0.37\textwidth]{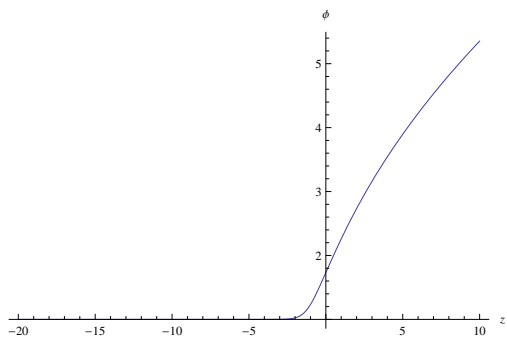}
\end{center}
\caption{Soliton scalar field configuration with $\xi=\upsilon=1$. We choose the coordinate such that potential varies along the $z$ direction.}
\label{config}
\end{figure}

In the flat space time, the scalar field configuration can be fixed by minimizing the energy of topological defect, which gives us 
\begin{eqnarray}
\frac{\partial\phi}{\partial z}=\sqrt{2V}~,
\end{eqnarray}
where we have chosen our coordinate such that potential varies along $z$ direction. Numerically the scalar field configuration can be solved out  as FIG.\ref{config}. Analytically, for large $\phi/\upsilon>>1$, we have the following approximation,
\begin{eqnarray}
\phi\propto z^{1/2}.
\end{eqnarray}


Following the argument in \cite{Vilenkin:1994pv}, suppose that our early universe start from a quantum era, with scalar field dispersion $\langle\phi^2\rangle>\upsilon^2$. In this epoch, scalar field doesn't feel the bump (or the well) in the potential. As universe expands, the scalar field dispersion goes down, and our universe was splited into two parts, one trapped in the well, the other run away to the infinite. 

A necessary condition for inflation to happen in the core of topological defect is that the thickness of the defect must be larger than the Hubble radius. The thickness $L$ of defect could be determined by the balance of gradient and potential energy, 
\begin{eqnarray}
\left(\frac{\sqrt{3}\upsilon}{L}\right)^2\sim V~.%\left(\phi=\sqrt{6}\Lambda\right)~.
\end{eqnarray}
$L>H^{-1}$ gives rise to such necessary condition for topological inflation, 
\begin{eqnarray}
\upsilon> M_p,
\end{eqnarray}
where the Freedman equation $3M_p^2H^2\simeq V$ has been used.

The slow roll parameters read
\begin{eqnarray}
%\partial V/\partial\phi&=&\xi  \left(-\frac{6 \Lambda ^{10}}{\phi ^7}+\frac{4 \Lambda ^8}{\phi ^5}-\frac{\Lambda ^6}{2 \phi ^3}\right)~,\nonumber\\
%\partial\partial V/\partial\phi^2&=&\xi  \left(\frac{42 \Lambda ^{10}}{\phi ^8}-\frac{20 \Lambda ^8}{\phi ^6}+\frac{3 \Lambda ^6}{2 \phi ^4}\right)\nonumber\\
\epsilon&=&\frac{2 M_p^2 \left(\phi ^2-3 \upsilon ^2\right)^2}{\phi^2\left(\phi ^2-\upsilon ^2 \right)^2},\nonumber\\
\eta&=&\frac{2 M_p^2 \left(21 \upsilon ^4-20 \upsilon ^2 \phi ^2+3 \phi ^4\right)}{\phi^2\left(\phi ^2-\upsilon ^2  \right)^2}~.
\end{eqnarray}
By requiring that 
\begin{eqnarray}
&&r=16\epsilon=0.16,~~n_s-1=2(\eta-3\epsilon)=-0.04,\nonumber\\
&&\frac{V}{\epsilon M_p^4}\sim 10^{-10},
\end{eqnarray}
we get 
\begin{eqnarray}
\phi\simeq 2.4\upsilon,~~~\upsilon\simeq3.5M_p,~~~\xi\sim10^{-11}.
\end{eqnarray}
At the end of inflation, the reheating could be triggered by a water fall scalar field, just like what people did in the  Hybrid inflation \cite{Linde:1993cn}\cite{Sasaki:2008uc}, and we are not going to discuss the detail of reheating here.

\section{Topological inflation with no-well potential}

\begin{figure}
\begin{center}
\includegraphics[width=0.37\textwidth]{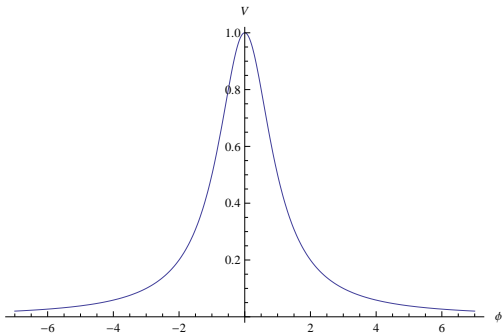}
\end{center}
\caption{Schematic plot of no-well scalar potential with $\xi=\upsilon=1$.}
\label{no1}
\end{figure}

\begin{figure}
\begin{center}
\includegraphics[width=0.37\textwidth]{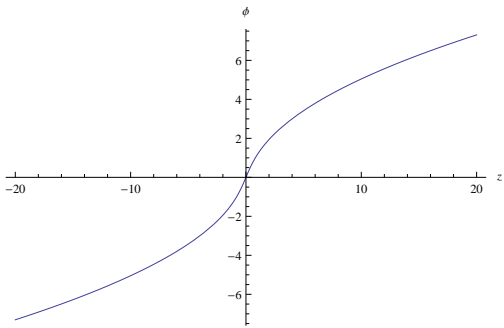}
\end{center}
\caption{Scalar field configuration for a no-well defect with $\xi=\upsilon=1$. }
\label{no2}
\end{figure}

Let us consider our last example, a no-well potential
\begin{eqnarray}
V(\phi)=\frac{\xi\upsilon^4}{1+\phi^2\upsilon^{-2}}~.
\end{eqnarray}
 Such type of potential was used to realized a matter bounce scenario in \cite{Lin:2010pf} (see \cite{Sakai:1998rg} for the 1st example of no-well topological inflation). The bump in this potential separates two domains, $\phi\to-\infty$ and $\phi\to+\infty$, see FIG.\ref{no1}. We numerically work out the scalar field configuration in flat space time. The result is shown in  FIG.\ref{no2}.

The necessary condition $L>H^{-1}$ yields
\begin{eqnarray}
\upsilon\gtrsim M_p~.
\end{eqnarray}

The slow roll parameters can be calculated as 
\begin{eqnarray}
\epsilon&=&\frac{2 \phi ^2 M_p^2}{\left(\upsilon^2+\phi ^2\right)^2}~,\nonumber\\
\eta&=&-\frac{2 \left(\upsilon^2-3 \phi ^2\right) M_p^2}{\left(\upsilon^2+\phi ^2\right)^2}~,\nonumber\\
\end{eqnarray}

By requiring that 
\begin{eqnarray}\label{17}
&&r=16\epsilon=0.16,~~n_s-1=2(\eta-3\epsilon)=-0.04,\nonumber\\
&&\frac{V}{\epsilon M_p^4}\sim 10^{-10},
\end{eqnarray}
we got 
\begin{eqnarray}
\phi\simeq0.7\upsilon,~\upsilon\simeq6.7M_p~, ~\xi\sim 10^{-16}~.
\end{eqnarray}

At the end of inflation, like the case of single well potential, reheating also needs to be triggered by a water fall scalar field. In this case, the end of inflation completely depends on the way inflaton couples to the water fall scalar,  and the shape of water fall scalar potential as well. Thus we can not extract any constraint from the requirement of e-folding number. That is the reason why we didn't consider the constraint from e-folding number in the above calculations.  The detailed study on the reheating is already beyond the scope of current paper, which will be covered in our future study. 

\section{Conlusion and Discussion}
In this paper, we check several topological inflationary models. We first classify the topological inflation models into three catagories,  double-well potential, single-well potential, and no-well potential inflation. 

In the case of double-well potential, we took the Higgs type of potential and the axion-like potential as examples. To be consistent with the observational constraints of red spectral tilt and large tensor-to-scalar ratio, we found that inflation of our Hubble volume should happen near the edge of solitons. 

Then we check second type of topological inflation, with only one single well. We took a potential motivated from extra-dimensional compactification as an example. With proper choice of parameters, such a model could be consistent with observations.

For the 3rd type of topological inflation, without a well in the potential, we also found that this model could be consistent with observations. 

~~~~~~~~~~~~~

\begin{acknowledgments}
The authors would like to thank Shinji Mukohyama, Yi Wang, Ryo Namba, Tomohiro Fujita, and Keisuke Harigaya for the usefull discussion. 

This work is supported by World Premier International Research Center Initiative (WPI), MEXT, Japan.

%Their contribution is helpful and important for the author to finish this work.
\end{acknowledgments}

\end{document}